\documentclass[aps,twocolumn,showpacs,amsmath,amssymb,superscriptaddress]{revtex4-1}

\usepackage{graphicx}
\usepackage{color}
\usepackage{subfigure}
\usepackage{dcolumn}
\usepackage{bm}
\usepackage{epsf}
\usepackage{amsmath,amstext,amssymb}
\usepackage{enumitem}
\usepackage{hyperref}
\usepackage[ansinew]{inputenc}
\usepackage{braket}
\usepackage{bbold}
\usepackage{siunitx}
\PassOptionsToPackage{numbers,sort&compress}{natbib}

\makeatletter
\g@addto@macro\normalsize{%
	\setlength\abovedisplayskip{4pt}
	\setlength\belowdisplayskip{4pt}
	\setlength\abovedisplayshortskip{4pt}
	\setlength\belowdisplayshortskip{4pt}
}
\makeatother

\newcommand{\be}{\begin{equation}}
\newcommand{\ee}{\end{equation}}
\newcommand{\bea}{\begin{eqnarray}}
\newcommand{\eea}{\end{eqnarray}}

\begin{document}
	
	\title{Scalable creation of long-lived multipartite entanglement}

	\author{H.~Kaufmann}
	\author{T.~Ruster}
	\author{C.~T.~Schmiegelow}\thanks{Present address: LIAF -  Laboratorio de Iones
		y Atomos Frios, Departamento de Fisica \& Instituto de Fisica de Buenos Aires,
		1428 Buenos Aires, Argentina}
	\author{M.~A.~Luda}\thanks{Present address: DEILAP, CITEDEF \& CONICET, J.B. de
		La Salle 4397, 1603 Villa Martelli, Buenos Aires, Argentina}
	\author{V.~Kaushal}
	\author{J.~Schulz}
	\author{D.~von Lindenfels}
	\author{F.~Schmidt-Kaler}
	\author{U.~G.~Poschinger}\email{poschin@uni-mainz.de}
	
	\affiliation{Institut f\"ur Physik, Universit\"at Mainz, Staudingerweg 7, 55128
		Mainz, Germany}

	\begin{abstract}
We demonstrate the deterministic generation of multipartite entanglement based on scalable methods. Four qubits are encoded in $^{40}$Ca$^+$, stored in a micro-structured segmented Paul trap. These qubits are sequentially entangled by laser-driven pairwise gate operations. Between these, the qubit register is dynamically reconfigured via ion shuttling operations, where ion crystals are separated and merged, and ions are moved in and out of a fixed laser interaction zone. A sequence consisting of three pairwise entangling gates yields a four-ion GHZ state $\ket{\psi}=\tfrac{1}{\sqrt{2}}\left(\ket{0000}+\ket{1111}\right)$, and full quantum state tomography reveals a Bell state fidelity of 94.4(3)\%. We analyze the decoherence of this state and employ dynamic decoupling on the spatially distributed constituents to maintain 69(5)\% coherence at a storage time of 1.1~seconds.		
	\end{abstract}
	
	
	\maketitle

The key challenge for the realization of quantum information processing devices which actually outperform classical information technology lies in the scaling to a sufficient complexity, while maintaining high operational fidelities. With trapped ions and superconducting circuits being the leading candidates for scalable high-fidelity quantum computing (QC) platforms, few-qubit architectures have been realized \cite{Kelly2015, SCHINDLER2013}, and elementary quantum algorithms \cite{MONZ2016,Debnath2016} as well as fundamental building blocks for quantum error correction \cite{CHOW2014,NIGG2014} have been demonstrated. For trapped ions, a possible route to scalability was opened up  with the seminal proposal of the \textit{quantum CCD} \cite{KIELPINSKI2002}, where ions are stored in segmented, micro-chip-based radiofrequency traps \cite{SEIDELIN2006,HENSINGER2006} and shuttled between distinct trap sites in order to realize quantum logic operations on selected subgroups of qubits \cite{ROWE2002,BOWLER2012,WALTHER2012,RUSTER2014,FALLEK2016}. Based on these methods, a complete methods set for QC \cite{HOME2009} and a fully programmable two-qubit quantum processor \cite{HANNEKE2010} have been shown. It is rather likely that any trapped-ion based large-scale QC architecture \cite{LEKITSCH2015,MONROE2014,BRANDL2017} will involve ion shuttling operations.\\
As a benchmark for quantum information processing capabilities, the generation and properties of multipartite entangled states have been studied intensively. On the one hand, generating and maintaining such states lies at the heart of quantum computing, on the other hand large multipartite entangled states represent a resource for the measurement-based approach to QC \cite{RAUSSENDORF2001, LANYON2013}. The first generation of a four-particle Greenberger-Horne-Zeilinger (GHZ) states has been accomplished at a state fidelity of 57\% by the NIST group \cite{Sackett2000}, while eight-qubit W-states at 76\% fidelity have been created later by the Innsbruck group \cite{HAEFFNER2005}. Furthermore, GHZ states of up to 14 trapped ions have been created \cite{Blatt14Qubit}, and it has been shown that these states are rather fragile in the presence of correlated magnetic field noise. While large-scale entanglement of thousands of optical modes \cite{Yokoyama2013} or atoms \cite{McConnell2015} has also been demonstrated, QC generally requires \textit{deterministic} entanglement generation with capabilities for storage and \textit{individual} manipulation and readout of the qubits.\\
In this work, we demonstrate the scalable generation of GHZ states of up to four trapped ions. Our method is based on single-qubit rotations, pairwise two-qubit entangling gates and shuttling operations. In analogy to arithmetic-logic-units (ALU) in the von-Neumann computer architecture \cite{BRANDL2017}, the computational gate operations are driven by laser beams which are directed to one fixed trap site, the laser interaction zone (LIZ). By shuttling only the required ions to this trap site, crosstalk is strongly suppressed as compared to static ion-crystal registers, where different ions are only spaced by a few microns. This is particularly beneficial, as memory ions, which are not to be affected by gate operations, are stored several hundreds of microns away from the LIZ. We also demonstrate that GHZ coherence can be maintained over long storage times by dynamical decoupling on the distributed components. Here, the constituent ions are kept in pairs and shuttled repeatedly into the LIZ, where the decoupling rotations take place. \\
\begin{figure}[h!tp]\begin{center}
		\includegraphics[width=0.47\textwidth]{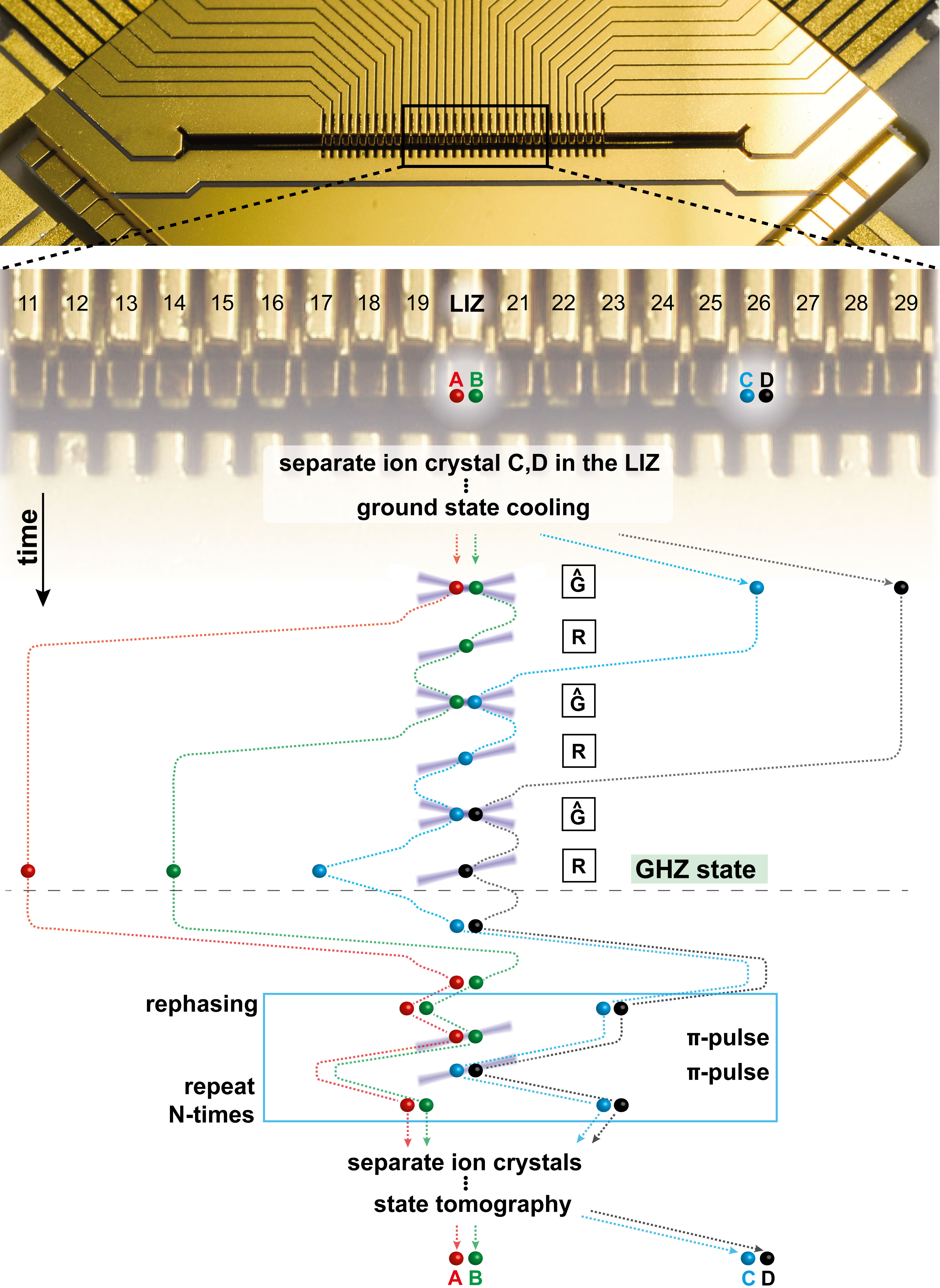}
		\caption{Experimental shuttling sequence for the creation, storage and analysis of a four-ion GHZ state. The state creation takes place above the dashed line. Laser driven quantum operations, i.e. entangling gates $\hat{G}$ or single qubit rotations $R$ are carried out in the laser interaction zone (LIZ). After the GHZ state is created, it can be stored for hundreds of ms by dynamical decoupling. The ions are not stored in the LIZ to prevent depolarization from residual light near the cycling transition.}
		\label{fig:shuttleSeq}
\end{center}\end{figure}
The centerpiece of our quantum processor is a segmented miniaturized Paul trap -- a photograph is shown in figure \ref{fig:shuttleSeq}. The design is similar to the one reported in Ref.~\cite{SCHULZ2008}. Quantum information is encoded in the Zeeman sublevels of the ground state of trapped $^{40}$Ca$^+$ ions $\ket{0}\equiv\ket{\downarrow}\equiv\ket{S_{1/2},m_J=-\tfrac{1}{2}}$ and $\ket{1}\equiv\ket{\uparrow}\equiv\ket{S_{1/2},m_J=+\tfrac{1}{2}}$. The Zeeman sublevels are separated by $2\pi\times$\SI{10.5}{\mega\hertz} by a magnetic field, which is generated by permanent magnets. The trap is placed in a $\mu$-metal enclosure for shielding of fluctuating ambient magnetic fields, and all experiments are synchronized to the ac line frequency. This yields a Ramsey coherence time of around \SI{300}{\milli\second} \cite{RusterLongLived2016} for a single qubit.

Laser light is employed for Doppler cooling (397~nm), qubit state initialization and qubit readout (729~nm), see Ref.~\cite{SWAPGATE}. For a fluorescence detection time of \SI{1.2}{\milli\second}, we achieve a measurement fidelity of 99.92\% for a single static ion. This fidelity is impaired by around 0.2\% in the four-ion sequence by the limited lifetime of the metastable $D_{5/2}$ state, since shelving is first applied to all ions, before fluorescence detection is carried out.

A separate laser source near 397~nm is used to manipulate the qubits via stimulated Raman transitions at a detuning from resonance of about $-2\pi\times$\SI{290}{\giga\hertz}. To cool the ions close to the motional ground state, a pair of orthogonally propagating beams is employed. The difference wave vector of the beams couples only to the radial secular modes, as it is oriented orthogonally to the trap axis. Therefore, the laser-ion interaction is insensitive to axial ion motion, excited by shuttling operations or anomalous heating. A second pair of identically aligned beams is used for pairwise qubit entanglement via a geometric phase gate \cite{LEIBFRIED2003A}.\\

In our setup, each laser beam is directed at the LIZ, and all operations are robust with respect to excitation of the axial ion motion: the entangling gate (driven on a radial secular mode), single qubit rotations (driven by co-propagating laser beams), electron shelving by 729~nm light (laser beam directed perpendicular to trap axis). Typical secular trap frequencies are: $\omega_{x,y,z}$= $2\pi\times\{1.5, 4.1, 4.9\}$ \SI{}{\mega\hertz}, where $x$ denotes the trap axis. Ions are shuttled along this direction by applying time-dependent waveforms to the dc electrodes of the linear Paul trap \cite{WALTHER2012,RUSTER2014}. To ensure a high entangling gate fidelity on a radial secular mode, the amplitude of the $2\pi\times$\SI{33}{\mega\hertz} rf trap drive is actively stabilized at a set-point of about \SI{160}{\volt}. We find the long-term relative frequency stability to be about 5~ppm. Anomalous heating rates on the radial secular modes range between 3(1) and 20(1) quanta per second, presumably limited by technical noise. The combination of low heating rates and long spin coherence times allows for concatenated quantum gate operations. \\


The experimental toolbox involves a set of techniques, which we describe in the following. Since the entangling gate requires the ions to be close to the ground state of motion, we apply resolved sideband cooling to all ions. A single cooling pulse takes on average \SI{15}{\micro\second}. We either cool single ions or pairs of two ions and cool all radial modes with 40 pulses per mode.\\

A single qubit $\pi$-rotation is realized by a \SI{10}{\micro\second} pulse. Using randomized benchmarking \cite{Knill2008}, we determine an error per computational gate as low as $5.1(2)\times 10^{-5}$. At the given Raman detuning and Rabi frequency, we find that the error is predominantly caused by residual photon scattering. Single qubit rotations can be performed on two ions simultaneously -- we calibrate the relative imbalance in terms of the time required for a $\pi$ rotation to better than $1.5 \times 10^{-5}$.\\ 

\begin{figure}[h!tp]\begin{center}
			\includegraphics[width=0.50\textwidth]{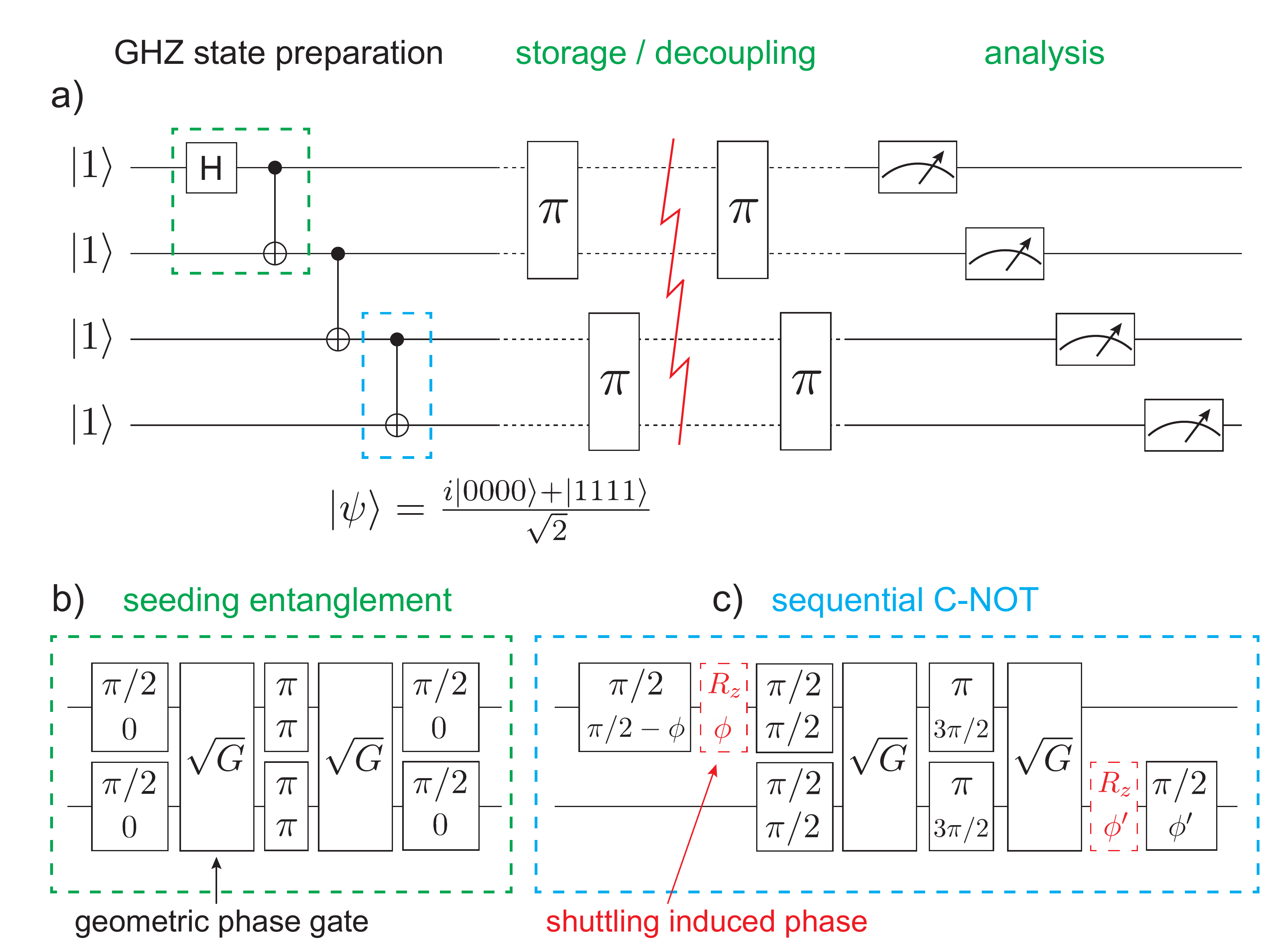}
			\caption{(a) Quantum circuit for the creation, storage and analysis of a four-ion GHZ state . The state is generated by three subsequent entangling gates. Evenly spaced dynamical decoupling $\pi$-pulses are employed to achieve long coherence times. followed by state tomography  on the individual ions. The green (b) and blue (c) dashed boxes show how the respective entangling gates are realized in the experiment -- they correspond to the unitary operators $\hat{U}_1$ and $\hat{U}_2$, from Eq. \ref{eq:unitary} . The black numbers for the single qubit operations represent the laser pulse areas, whereas gray numbers indicate the phase of the respective pulse. Red boxes represent an induced qubit phase which arises from the shuttling operations, see text. }
			\label{fig:circuit}
	\end{center}\end{figure}

The entangling gate is driven by spin-dependent optical dipole forces \cite{LEIBFRIED2003A}, generating the unitary $\hat{G}=\text{diag}(1,i,i,1)$.
At a detuning of $2\pi\times$~\SI{25}{\kilo\hertz} from the higher frequency radial center-of-mass mode, the gate operation takes \SI{100}{\micro\second}. We achieve a fidelity of 99.0(4)~\%. As for the qubit rotations, the fidelity is limited by off-resonant photon scattering. 
The gate is used, in conjunction with single qubit rotations, to generate the unitary $\hat{U}_1$ for entanglement seeding and the CNOT equivalent unitary $\hat{U}_2$:
		\begin{equation}
	\hat{U}_1=\frac{e^{i\pi/4}}{\sqrt{2}}
	\begin{pmatrix}
		1 & 0 & 0 & i \\ 
		0 & 1 & -i & 0 \\ 
		0 & -i & 1 & 0 \\ 
		i & 0 & 0 & 1
	\end{pmatrix}
	, \quad
	\hat{U}_2=
	\begin{pmatrix}
		0 & 1 & 0 & 0 \\ 
		-i & 0 & 0 & 0 \\ 
		0 & 0 & i & 0 \\ 
		0 & 0 & 0 & 1
	\end{pmatrix},\label{eq:unitary}
\end{equation}
see also Fig. \ref{fig:circuit}. The sequential CNOT $\hat{U}_2$ requires shuttling of a single ion to the LIZ for a single qubit rotation, a subsequent recombination of two ions in the LIZ for an entangling gate, followed by a separation operation and a final transport of a single ion to the LIZ for a single qubit rotation.\\

Ion shuttling along the trap axis is performed by concatenated segment-to-segment transports of \SI{200}{\micro\meter}, where each operation takes \SI{30}{\micro\second}. We estimate the motional excitation on the radial gate mode to be below 0.01 phonons per transport. Separation of a two-ion crystal is realized by increasing the voltage on the trapping segment and lowering the voltage on the adjacent segments \cite{KaufmannSplitTheory,RUSTER2014} within \SI{160}{\micro\second}. The operation and its reversed counterpart -- the recombination of two individual ions -- are only carried out in the LIZ as it requires careful calibration of the electrode voltages via precise spectroscopic measurements of motional sideband frequencies. Separation and recombination operations lead to a motional excitation of about 0.05 phonons on the gate mode.\\

Since the ions are shuttled along the trap axis while in a spin superposition state, they accumulate a phase $\phi$ due to the inhomogeneous magnetic field along the trap axis \cite{Ruster2017}. In total four of these phases need to be considered in the quantum gate sequence. We find each of them to be constant over time and to be $\phi< \pm 0.6$ rad. The phase is compensated for by adjusting the phase of an adjacent single qubit rotation, see Fig. \ref{fig:circuit} c).\\
 After the application of state tomography laser pulses, each ion is shuttled to the LIZ for the population state transfer $\ket{\uparrow} \leftrightarrow \ket{D_{5/2}}$ via electron shelving. The ions are again individually shuttled to the LIZ, where state dependent fluorescence is observed. All qubits are shelved \textit{before} fluorescence detection, to avoid depolarization of a remotely stored qubit from scattered light near 397 nm.
	
	
In order to generate multi-partite entanglement, we employ a sequence of two-ion entangling gates and single qubit operations to create a GHZ state $\ket{\Psi} = 1/\sqrt{2}~(i\ket{0000}+\ket{1111})$. The experimental sequence, including the shuttling operations, is sketched in Fig. \ref{fig:shuttleSeq}. It is comprised of five blocks: An initial cooling block prepares the ions close to the ground state of motion of the radial secular modes, which is crucial for high-fidelity gate operations in the subsequent quantum logic block. Once the GHZ state is prepared, the coherence can be maintained through an optional rephasing block. To analyze the final quantum state, full quantum state tomography is performed in the analysis block. A final block contains an additional magnetic field tracking measurement. Furthermore, the ions are repositioned to their initial positions in order to enable the next repetition of the entire sequence. In the following, we describe these operational blocks in detail.\\
Initially, the ion pair $A,B$ is stored at electrode 20 in the LIZ and the ion pair $C,D$ is stored at electrode 26. The ion pair $A,B$ is then shuttled to electrode 14 and the pair $C,D$ is shuttled to the LIZ for a separation operation which transfers ion $C$ to electrode 19 and ion $D$ to electrode 21. The ion pair $A,B$ and the single ions $C$ and $D$ are transported to the LIZ for ground state cooling and spin initialization to $\ket{\psi}=\ket{\uparrow\uparrow\uparrow\uparrow}\equiv\ket{1111}$. Subsequently, we perform quantum logic operations to generate the maximally entangled GHZ state, given by the quantum circuit shown in Fig. \ref{fig:circuit}. 
	The circuit is comprised of one and two qubit quantum logic operations, which are carried out in the LIZ. Initially the quantum state is set to $\ket{\psi}=\ket{1111}$. Application of the unitary  $\hat{U}_1$ on the ion pair $A,B$ in the LIZ generates the state $\ket{\psi}=(i\ket{0011}+\ket{1111})/\sqrt{2}$, where ions A and B are entangled. The entanglement is extended to all qubits by subsequent application of the C-NOT unitary $\hat{U}_2$ on qubits B,C and C,D. This leads to the final state $\ket{\psi}=(i\ket{0000}+\ket{1111})/\sqrt{2}$, where the constituent qubits are distributed over a macroscopic distance of \SI{1.8}{\milli\meter}.\\	
 To assess the fidelity of the generated state, we perform full quantum state tomography by subsequent shuttling of each of the ions to the LIZ. Here, for each ion one of the analysis rotations \{$\mathbb{1},R_X(\pi/2),R_Y(\pi/2)$\} is driven, in order to measure the operators \{$\sigma_z,\sigma_y,\sigma_x$\}. This amounts to 81 different measurement settings for the four qubits. The analysis pulses are not corrected for additional phase accumulations due to the magnetic field inhomogeneity on individual ions, as we only assess the gauge-invariant fidelity of the GHZ state irrespective of an additional phase.\\

\begin{figure}[h!tp]\begin{center}
			
			\includegraphics[width=0.50\textwidth]{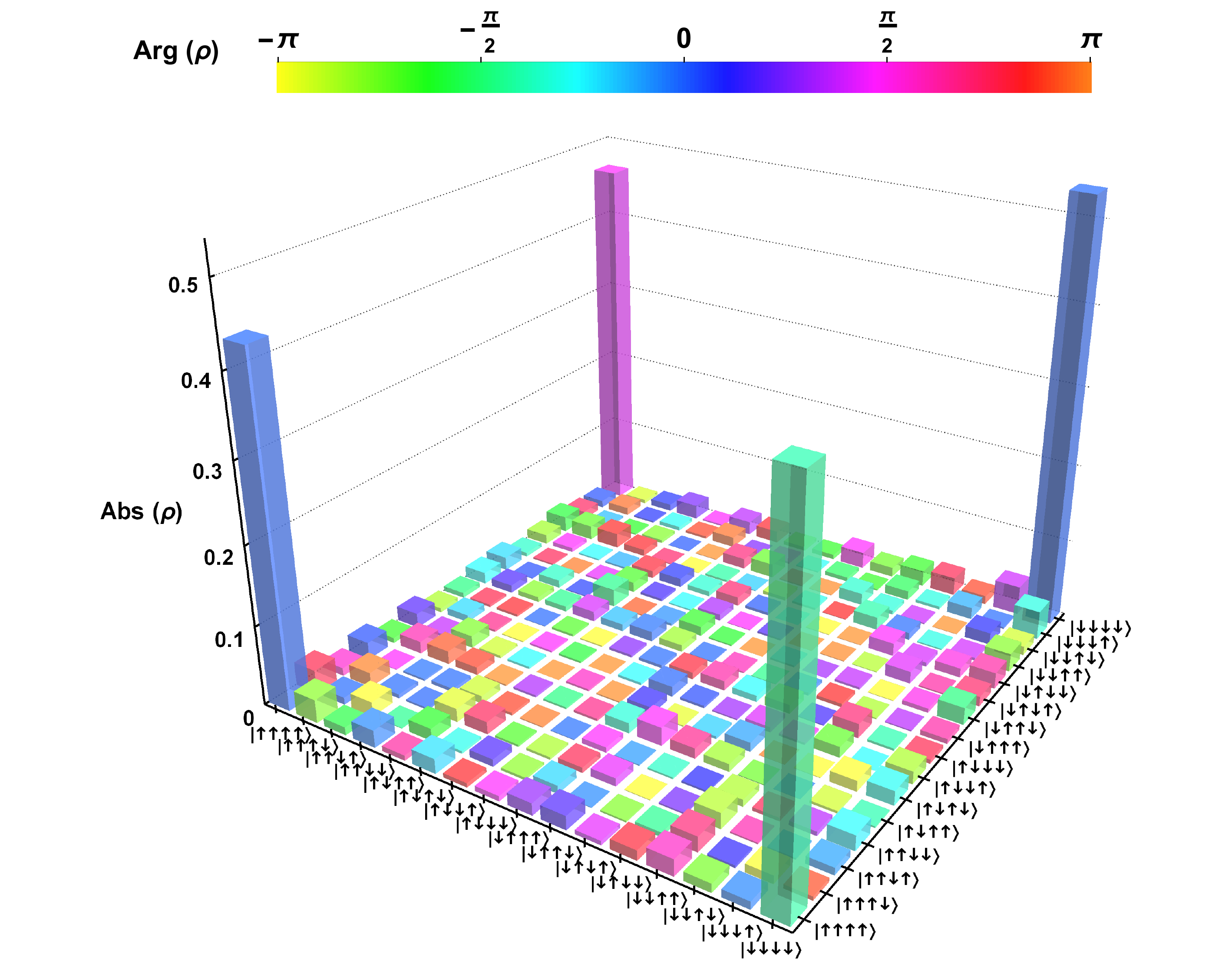}
			\caption{Reconstructed density matrix of a maximally entangled four-ion GHZ
				state with correction for readout errors. Linear reconstruction with correction for SPAM errors yields a state fidelity of $\mathcal{F}=94.38$\%.
			}
			\label{fig:densityMat}
	\end{center}\end{figure}

After readout of all four ions, we carry out the magnetic field tracking block. A single ion is used to measure the deviation of the carrier frequency from the Zeeman splitting in a Ramsey experiment, i.e. the absolute magnetic field. Specifically, ion A is initialized to a superposition state with a $\pi/2$-pulse, which is followed by a \SI{5}{\milli\second} interrogation time and a final $\pi/2$-pulse with a phase of \SI{0}{\degree} or \SI{90}{\degree}. From the difference of these two measurements we calculate the deviation from the carrier frequency, which is updated for subsequent measurement cycles and thus compensates for small absolute magnetic field drifts. Finally, the ions are
	merged into pairs and the entire sequence is repeated. The obtained density
	matrix is displayed in Fig. \ref{fig:densityMat}.
	
In order to quantitatively characterize the fidelity of the prepared state, we reconstruct density matrices $\hat{\rho}$ from measurement data and compute the state fidelity
\begin{equation}
\mathcal{F}(\hat{\rho})=\underset{\theta}{\text{max}}\;\bra{\Psi(\theta)}\hat{\rho}\ket{\Psi(\theta)}
\end{equation}
with respect to a GHZ state of arbitrary relative phase $\theta$:
\begin{equation}
\ket{\Psi(\theta)}=\tfrac{1}{\sqrt{2}}\left(\ket{0000}+e^{i\theta}\ket{1111}\right).
\end{equation}

We first perform linear inversion of the measurement data, which consists of 4$\times$50900 measurements in total. This yields a density matrix, from which a fidelity of $\mathcal{F}=$92.60\% is extracted. As a density matrix obtained from linear inversion can have negative eigenvalues due to statistical errors, the density matrix obtained from linear inversion can not be used for estimation of a confidence interval via parametric bootstrapping. We additionally perform a maximum-likelihood (ML) state reconstruction \cite{REHACEK2007}. Using the physical density matrix from the ML reconstruction for parametric bootstrapping, we estimate a fidelity of $\mathcal{F}_{ML}=$92.50(37)\%, such that linear inversion and ML estimation yield consistent results.\\
By running the sequence without quantum logic operations, we determine the readout errors \cite{SWAPGATE}. State preparation and measurement (SPAM) errors are dominated by the limited lifetime of the metastable $D_{5/2}$ state of \SI{1.2}{\second}. For ion $A$, for instance, the time between shelving and fluorescence detection on the cycling transition is \SI{2.7}{\milli\second}, leading to an estimated decay-induced SPAM error of 0.2\%. An actual error of 0.4\% is measured for ion $A$, the remaining error is attributed to shuttling-induced motional excitation, which affects the shelving efficiency. This is either due to residual coupling of 729~nm shelving laser to axial motion, or due to residual radial excitation from shuttling. Including correction for SPAM errors, the fidelity obtained from linear inversion is $\mathcal{F}=$94.38\%, and ML estimation with parametric bootstrapping yields $\mathcal{F}_{ML}=$94.28(30)\%. Additionally, we perform statistical tests based on Hoeffding's tail inequality \cite{MORODER2013}, confirming that the measurement data is statistically consistent with the state described by the reconstructed density matrices.
The theoretical fidelity limit for our setting is 97\%, since the entangling gate is performed three times and features a fidelity of 99.0(4)\%. We attribute the discrepancy from this result to imperfect calibration of the individually calibrated entangling gates and the correction phases $\phi$, as well as the finite accuracy of the stabilized magnetic field tracking measurements. The entangling gates require individual calibration due to small motional excitation -- mainly from heating -- which results in a lower coupling to the driving field and thus requires a slightly increased duration of the entangling gate pulse.\\	
	The execution time for the creation of the GHZ state after sideband cooling is \SI{3.1}{\milli\second}, where 11\%  is used for quantum gates and the remainder is dedicated to shuttling operations. This illustrates that currently, the shuttling overhead dominates the time budget of the quantum CCD operation. However, a significant leeway for optimization of these operations via technological and methodological and parallelization improvements exists.

\begin{figure}[h!tp]\begin{center}
			\includegraphics[width=0.45\textwidth]{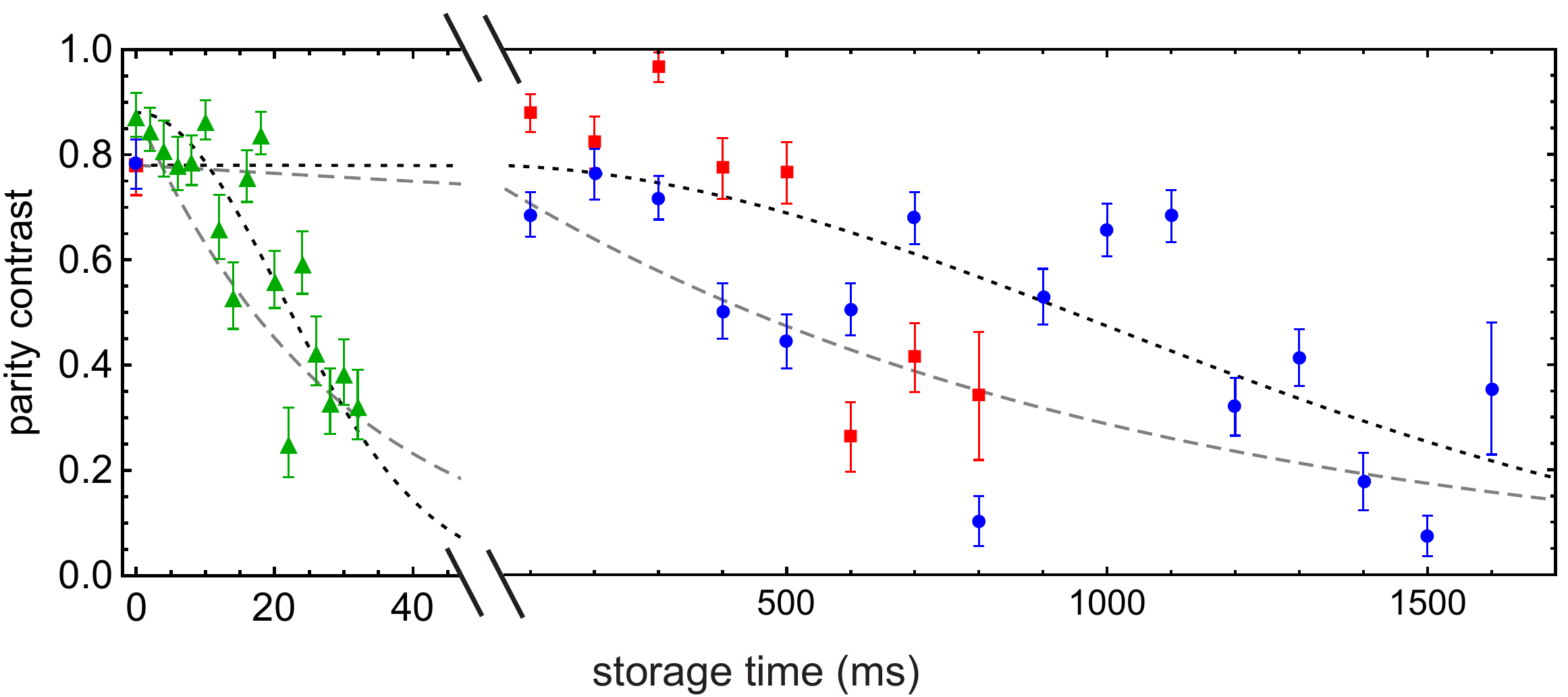}
			\caption{Preservation of the parity contrast of a four-ion GHZ state by dynamical decoupling. Green triangles for short times correspond to a measurement without the execution of the rephasing block. Blue circles  represent a measurement with $N_\pi = 15$ rephasing pulses on each qubit, while red squares correspond to  $N_\pi = 3$. The dotted black line represents a Gaussian decay, whereas the grey dashed line represents an exponential decay. The data shown is not corrected for SPAM errors.
			}
			\label{fig:rephaseResults}
	\end{center}\end{figure}

Magnetically sensitive multi-qubit GHZ states are prone to \textit{super-decoherence}, which is caused by correlated local magnetic field noise\cite{Blatt14Qubit}. We measure the coherence time of a four-ion GHZ state and employ a dynamical decoupling technique to achieve unprecedented long lifetimes. Here, repeated $\pi$-flips of the qubits serve to cancel the coupling of the qubits to an offset magnetic field \cite{Vandersypen2005}. Once the GHZ state is created, the four individually trapped ions are merged to ion pairs $A,B$ and $C,D$ to reduce the amount of shuttling operations in the rephasing block. After storing the ion pairs at segments 19 and 23, they are alternately shuttled to the LIZ and subjected to rephasing $\pi$-pulses. The pulses are evenly spaced within the storage time, thus only an odd number of pulses is suitable for decoupling. After the rephasing block, the ion pairs are separated into individually trapped ions for state analysis. We utilize a reduced measurement scheme, by measuring the operators $\{\sigma_x^{(A)}\sigma_x^{(B)}\sigma_x^{(C)}\sigma_x^{(D)}\}$ and $\{\sigma_x^{(A)}\sigma_x^{(B)}\sigma_x^{(C)}\sigma_y^{(D)}\}$. To infer the GHZ coherence of the state, each of the two operators is measured at least 200 times. The parity contrast and its statistical measurement error are determined from the measurement results via Bayesian parameter estimation \cite{Ruster2017}. The results of these measurements are shown in Fig. \ref{fig:rephaseResults}. 
	The coherence time of the GHZ state without rephasing pulses and without recombining the ions to pairs, is about \SI{20}{\milli\second}. By applying $N_\pi = 15$ rephasing pulses on each ion pair, we preserve the coherence for storage times exceeding one second. As can be seen from Fig. \ref{fig:rephaseResults}, the coherence decay is not described by a simple model. For several intermediate storage times the remaining coherence is lower, we attribute this to noise at frequencies that match the inverse time difference between subsequent decoupling pulses \cite{Kotler2011}, since the coherence can still be preserved by choosing a different number of rephasing pulses $N_\pi$. 
	
	
	In conclusion, we demonstrate the scalable creation of a maximally entangled four-qubit Greenberger-Horne-Zeilinger state with high fidelity, in a trapped-ion quantum processor. The state is generated by a sequence of single- and two-qubit gate operations, interleaved by shuttling operations that reconfigure the register. We employ dynamical decoupling to preserve the coherence of the sensitive entangled state for more than one second, which is an important feature for quantum error correction schemes \cite{NIGG2014}. We therefore demonstrate the \textit{deterministic} generation of multipartite entangled state, with its constituents distributed over a \textit{macroscopic range}, which persists for and unprecedented \textit{long storage time}.
	The combination of shuttling-insensitive quantum operations, low heating rates and long coherence times enables our programmable quantum processor to execute more advanced quantum algorithms with a larger number of qubits in the future. Technical and methodological improvements will allow for much faster shuttling operations and thus increase the operational speed.	In conjunction with feedback operations and fast physical two-ion SWAP gates - which have been realized in our trap with high fidelity \cite{SWAPGATE} - our setup will be extended to a freely programmable quantum computing platform.

	The research is based upon work supported by the Office of the Director of National Intelligence (ODNI), Intelligence Advanced Research Projects Activity (IARPA), via the U.S. Army Research Office grant W911NF-16-1-0070. The views and conclusions contained herein are those of the authors and should not be interpreted as necessarily representing the official policies or endorsements, either expressed or implied, of the ODNI, IARPA, or the U.S. Government. The U.S. Government is authorized to reproduce and distribute reprints for Governmental purposes notwithstanding any copyright annotation thereon. Any opinions, findings, and conclusions or recommendations expressed in this material are those of the author(s) and do not necessarily reflect the view of the U.S. Army Research Office. We acknowledge helpful discussions with Philipp Schindler and Alex Retzker, and financial support from the DFG through the DIP program (Grant No. SCHM 1049/7-1).
	

	\bibliographystyle{apsrev4-1}
	\bibliography{lit}
	
\end{document}